\documentclass[%
 aip,
jmp,%
sd,%
amsmath,amssymb,
preprint,%
onecolumn,%
fleqn,%
groupedaddress,%
tightenlines,%
]{revtex4-1}

\usepackage{graphicx}
\usepackage{dcolumn}
\usepackage{bm}

\usepackage{verbatim}
\usepackage{amsfonts}
\usepackage{amsmath}
\usepackage{amssymb}
\usepackage{amsthm}
\DeclareMathAlphabet{\mathpzc}{OT1}{pzc}{m}{it}
\newcommand{\ud}{\mathrm{d}}

\newcommand{\Tr}{\,\mathrm{Tr}}

\newcommand{\vol}{\mathrm{vol}}
\newcommand{\sgn}{\mathrm{sgn}}

\begin{document}


\title{Nonperturbative evaluation of the partition function for the real scalar quartic QFT on the Moyal plane at weak coupling}

\author{J. de Jong}
 \email{j.dejong@uni-muenster.de}
 \affiliation{Mathematisches Institut, Westf\"alische Wilhelms-Universit\"at M\"unster, Einsteinstra\ss{}e 62,
D-48149 M\"unster, Germany}
\author{R. Wulkenhaar}%
 \email{raimar@math.uni-muenster.de}
 \affiliation{Mathematisches Institut, Westf\"alische Wilhelms-Universit\"at M\"unster, Einsteinstra\ss{}e 62,
D-48149 M\"unster, Germany}%


\date{\today}

\begin{abstract}
The remarkable properties of the real scalar quartic quantum field theory on the Moyal plane in combination with its similarity to the Kontsevich model make the model's partition function an interesting object to study. However, direct evaluations is obstructed by the intertwining of the field's various modes. A factorization procedure to circumvent this problem is proposed and discussed here in the context of the real scalar quartic \textsc{qft} on the Moyal plane. This factorization consists of integrating against the asymptotic volume of the diagonal subpolytope of symmetric stochastic matrices. This volume has been determined to this end. Using this method the partition function for regime of weak coupling is computed. Using the same method it is as well possible to determine the partition function and free energy density for other regimes.
\end{abstract}

\maketitle

\section{Introduction\label{sec:intro}}
Since the first definition of the Wightman
axioms~\cite{wightman1,streater,wightman2} of quantum field theory it
has been attempted to find an example of a nontrivial \textsc{qft} in
four spacetime dimensions. Much progress has been booked and many
approaches have been tested, but up to this date all with limited
success.
\\

Partition functions over Hermitean matrices are a common tool for
two-dimensional quantum gravity. The graphs they generate are dual to
triangulations of surfaces. A particularly well-known example is the
Kontsevich model~\cite{kontsevich}, which was used to link such
theories to the intersection theory on the compactified moduli space
of complex curves.
\\

Such matrix models have a natural interpretation as a \textsc{qft} on
noncommutative Moyal space. An orthonormal basis of double indexed
functions exists\cite{GraciaBondia:1987kw} which maps the Moyal
product of real functions to products of Hermitean matrices. Using
this correspondence, scalar fields in a Euclidean \textsc{qft} 
on Moyal space may be expanded in Hermitean
matrices, where the size $N$ of the matrices is introduced as a
regulator. In this
way the Kontsevich model may be interpreted as a noncommutative 
\textsc{qft} in an even number of dimensions. This model was 
studied nonperturbatively by
Grosse and Steinacker~\cite{steinacker1, steinacker2, steinacker3}. A
quartic model for complex scalar fields was exactly
solved~\cite{langmann1} and found to be trivial.
\\

The quartic scalar \textsc{qft} for real fields was the natural model
to study after the cubic Kontsevich model. It may be thought of as a
generalization of the ordinary $\Phi^{4}$-theory in $d$ even
dimensions
\begin{align}
&\hspace{-5mm}S[\varphi]=\int \ud^{d} x\, \frac{1}{2}\varphi(x)\big(-\Delta+\Omega^{2}\Vert2\Theta^{-1}x\Vert^{2}+\mu^{2}\big)\varphi(x)+g\varphi^{\star 4}(x)\label{e:GW}\quad,
\end{align}
where the ordinary pointwise product is replaced by the Moyal product
with deformation matrix $\Theta$ and the propagator is supplemented
with a harmonic oscillator potential~\cite{wg0}. 
The harmonic oscillator potential ensures that the operator
$\big(-\Delta+\Omega^{2}\Vert2\Theta^{-1}x\Vert^{2}+\mu^{2}\big)$ 
in (\ref{e:GW}) has compact resolvent, which is necessary to deal with
the \textsc{uv}/\textsc{ir}-mixing problem~\cite{minwalla}.  
The action (\ref{e:GW}) is then
studied at the self-dual\cite{langmann2} point $\Omega=1$ which
gives rise to a trace under the above correspondence between Moyal
space and matrices.\\

A big difference with conventional $\varphi^{4}$-theory is the
vanishing of the $\beta$-function~\cite{disertori}. Usually, a
positive $\beta$-function indicates a Landau pole, so that the coupling
constant will diverge at a finite energy scale. To overcome this, the
coupling must be scaled to zero from the start, leading to a free
theory.\\

This quartic real scalar \textsc{qft} has been studied
intensively~\cite{wg1,wg2,wg3,wg4,wg5,wg6}. The Ward identies for this
model in combination with the Schwinger-Dyson equations yield a closed
equation for the $2$-point function in the limit of infinite
noncommutativity. In the Schwinger-Dyson methods used to study this
model, the renormalization is performed simultaneously with the limit
$N\rightarrow\infty$ that extends the Hermitean matrices back to the
full algebra. No direct evaluation of the partition function is
used. Motivated by success of the partition function approach 
on $\varphi^{\star 3}$ (Kontsevich) model,
it may be hoped that additional information about the $\varphi^{\star
  4}$ model may be obtained in this way.
\\

In Section \ref{sec:model} the real scalar quartic \textsc{qft} on
Moyal space will be formulated as a matrix model partition
function. The main obstacle to evaluation of this partition function
is the intertwining of the matrix' eigenvalues. To overcome this the
partition function will be factorized in Section \ref{sec:fact} using
the volume of a class of subpolytopes of symmetric stochastic
matrices. It is then possible to determine the partition functions for
the free theory and for weak coupling in Sections \ref{sec:noco} and
\ref{sec:weco} respectively. Simultaneoulsy, these computations are
tests of the factorization procedure.
\\
This article is a summary of the main results from the first author's Ph.D. thesis~\cite{jins2}.

\section{The matrix formulation of the model\label{sec:model}}
The partition function corresponding to (\ref{e:GW}) is
\begin{equation}
\mathpzc{Z}[J]=\int_{H_{N}(\mathbb{R})}\ud X\,\exp[\Tr\Big(-E X^{2}-g X^{4}+JX\Big)]\label{e:qmm1}\quad,
\end{equation}
where we assume naturally that $g>0$. The set of Hermitean $N\times
N$-matrices with real entries is denoted by $H_{N}(\mathbb{R})$. The
matrix $E$ corresponds to the Laplacian and is an unbounded
self-adjoint matrix with compact resolvent. It may be assumed diagonal
with the kinetic eigenvalues as entries.
\\

Evaluation of the partition function (\ref{e:qmm1}) is not straighforward. To keep grip on the process the model is studied at weak positive coupling, so that the result may be tested with the result for the free theory, which means that the coupling $g$ in (\ref{e:qmm1}) is set to zero. As will be explained in a moment, this will be done for $J=0$ too. The decomposition of the Hermitean matrix
\begin{equation*}
X=\sum_{k=1}^{N}X_{kk}^{(r)}+\sum_{1\leq k<l\leq N}X_{kl}^{(r)}+i X_{kl}^{(i)}+\sum_{1\leq n<m\leq N}X_{mn}^{(r)}-i X_{mn}^{(i)}
\end{equation*}
corresponds to the integration measure
\begin{equation*}
\ud X=\Big\{\prod_{1\leq k<l\leq N}\int_{-\infty}^{\infty}\ud X_{kl}^{(r)}\,\int_{-\infty}^{\infty}\ud X_{kl}^{(i)}\Big\}\times\Big\{\prod_{k=1}^{N}\int_{-\infty}^{\infty}\ud X_{kk}^{(r)}\Big\}\quad.
\end{equation*}
Here and in the rest of the paper the index notation $k<l$ will be used for $1\leq k<l\leq N$ and $\sum_{k}$ will be used for $\sum_{k=1}^{N}$.\\

In these components it follows that the trace of the square of such matrices is given by
\begin{equation*}
\Tr EX^{2} = \sum_{k=1}^{N}E_{kk}\big[(X_{kk}^{(r)})^{2}+\sum_{l=1}^{k-1}(X_{lk}^{(r)})^{2}+(X_{lk}^{(i)})^{2}+\sum_{l=k+1}^{N}(X_{kl}^{(r)})^{2}+(X_{kl}^{(i)})^{2}\big]\,.
\end{equation*}
This implies that the free partition function is given by
\begin{equation}
\mathpzc{Z}[0]=\Big\{\prod_{k=1}^{N}\sqrt{\frac{\pi}{e_{k}}}\,\Big\}\cdot\Big\{\prod_{1\leq k<l\leq N}\frac{\pi}{e_{k}+e_{l}}\Big\}\quad.\label{e:zerocoup}
\end{equation}

The matrix $J$ in (\ref{e:qmm1}) lies around zero and represents a
source term. However, the part of the $2$-point function that is the
most difficult to access from the Schwinger-Dyson equations \cite{wg4} should be extractable from the partition function for the vacuum sector, i.e. $J=0$. This simplifies the technical challenges of the model considerably. For example, the Harish-Chandra-Itzykson-Zuber integral~\cite{harish-chandra,itzykson-zuber, zinnP} can be applied to eliminate the matrix integral, so that
\begin{equation}
\mathpzc{Z}[0]=\frac{1}{N!}\frac{(-\pi)^{\binom{N}{2}}}{\Delta(e_{1},\ldots,e_{N})}\int_{\mathbb{R}^{N}}\ud \vec{\lambda}\,\frac{\Delta(\lambda_{1},\ldots,\lambda_{N})^{2}}{\Delta(\lambda_{1}^{2},\ldots,\lambda_{N}^{2})}\exp[-\sum_{j}g\lambda_{j}^{4}]\det_{k,l}\big(e^{-e_{k}\lambda_{l}^{2}}\big)\label{e:qmm2}\quad,
\end{equation}
where
\begin{equation*}
\Delta(\lambda_{1},\ldots,\lambda_{N})=\prod_{1\leq k<l\leq N}(\lambda_{l}-\lambda_{k})
\end{equation*}
is the Vandermonde determinant of these eigenvalues. Rewriting (\ref{e:qmm2}) yields
\begin{align}
&\hspace{-8mm}\mathpzc{Z}[0]=\frac{(-\pi)^{\binom{N}{2}}}{(N!)\cdot\Delta(e_{1},\ldots,e_{N})}\!\sum_{\sigma\in\mathcal{S}_{N}}\!\!\sgn(\sigma)\!\int_{\mathbb{R}^{N}}\!\!\!\!\!\ud^{N}\vec{\lambda}\,\big[\!\!\!\prod_{1\leq k<l\leq N}\frac{\lambda_{l}-\lambda_{k}}{\lambda_{l}+\lambda_{k}}\big]e^{-\sum_{j}\lambda_{j}^{2}e_{\sigma(j)}+g\lambda_{j}^{4}}\nonumber\\
&=\frac{(-\pi)^{\binom{N}{2}}}{\Delta(e_{1},\ldots,e_{N})}\int_{\mathbb{R}^{N}}\!\!\!\ud^{N}\vec{\lambda}\,\big[\!\!\!\prod_{1\leq k<l\leq N}\frac{\lambda_{l}-\lambda_{k}}{\lambda_{l}+\lambda_{k}}\big]e^{-\sum_{j}\lambda_{j}^{2}e_{j}+g\lambda_{j}^{4}}\quad.\label{e:ZJ0c}
\end{align}
In the last step we renamed in each term of the sum over the permutation group the integration variables $k\mapsto \sigma(k)$. This maps the Vandermonde determinant in the numerator
\begin{equation*}
\prod_{1\leq k<l\leq N}(\lambda_{l}-\lambda_{k})\mapsto \prod_{1\leq k<l\leq N}(\lambda_{\sigma(l)}-\lambda_{\sigma(k)})=\sgn(\sigma)\prod_{1\leq k<l\leq N}(\lambda_{l}-\lambda_{k})\quad.
\end{equation*}
This shows that the divergences, $\lambda_{m}=-\lambda_{n}$, that
appears in (\ref{e:ZJ0c}) are fictitious. They are not present in
(\ref{e:qmm2}).
\\

It is not a big restriction to study the vacuum sector of the matrix
model. Alternatively, the extension
$\exp[-\Tr(X^{4}+(E+J)\cdot(X^{2}+X\kappa))]$ of the matrix action may
be considered. The kinetic eigenvalues $e_{j}$ in the vacuum sector
are then replaced by the eigenvalues of the Hermitean matrix
$E+J$. The consequence of the shift $X^{2}+X\kappa$ is that a linear
exponential factor $\exp[-\kappa\sum_{j}e_{j}\lambda_{j}]$ must be
added to (\ref{e:ZJ0c}). This action can be treated by the same
methods as the vacuum sector. The terms in the power series in
$\kappa$ and the entries $J_{kl}$ of this shifted partition function,
where the powers of $\kappa$ and $J$ are identical, compose the full
partition function. Successfully computing the vacuum sector of the
partition function demonstrates then that the partition function of
the full theory can be computed. This forms an extra argument to study
only the vacuum sector $\mathpzc{Z}[0]$ of the partition function.
\\

There is an obvious obstacle towards the integration of the partition
function (\ref{e:ZJ0c}) for large $N$. It is the intertwinement of the
eigenvalue integrals through the denominator. Performing the integral
over $\lambda_{1}$ would change the form of the integrand. This
implies that either an iterative integration scheme must be found that
fits these integrands or a way to reformulate the denominator in such
a way that the partition functions factorizes. The second approach is
pursued here.
\\

As noted above, the divergences at  $\lambda_{m}=-\lambda_{n}$ 
in (\ref{e:ZJ0c}) are fictitious. This means that the integration 
over $\lambda_n$ can exclude $\bigcup_{m>n} [-\lambda_m-\varepsilon, 
-\lambda_m+\varepsilon]$. Such principal value integrals are induced by 
\begin{align}
\frac{1}{\lambda_{k}+\lambda_{l}}
&\mapsto 
\frac{1}{2(\lambda_{k}+\lambda_{l}-\mathrm{i}\varepsilon)}
+\frac{1}{2(\lambda_{k}+\lambda_{l}+\mathrm{i}\varepsilon)}
\nonumber
\\
&=
+\frac{i}{2}
\int_{0}^{\infty}\ud u_{kl}\,e^{-iu_{kl}(\lambda_{k}+\lambda_{l}-i \varepsilon)}
-\frac{i}{2}
\int_{0}^{\infty}\ud u_{kl}\,e^{iu_{kl}(\lambda_{k}+\lambda_{l}+i \varepsilon)}\quad.
\label{e:ST}
\end{align}
Applying this strategy to the denominators in (\ref{e:ZJ0c}) shows
that 
\begin{align}
&\hspace{-5mm}\int_{\mathbb{R}^{N}}\ud\vec{\lambda}\,\prod_{k<l}\frac{\lambda_{l}-\lambda_{k}}{\lambda_{l}+\lambda_{k}}\nonumber\\
&=\int_{\mathbb{R}^{N}}\ud\vec{\lambda}\prod_{k<l}\Big\{\frac{i(\lambda_{l}\!-\!\lambda_{k})}{2}\int_{0}^{\infty}\!\!\ud u_{kl}\,e^{-iu_{kl}(\lambda_{k}+\lambda_{l}-i\varepsilon)}-\frac{i(\lambda_{l}\!-\!\lambda_{k})}{2}\int_{0}^{\infty}\!\!\ud u_{kl}\,e^{iu_{kl}(\lambda_{k}+\lambda_{l}+i\varepsilon)}\Big\}\nonumber\\
&=\int_{\mathbb{R}^{N}}\ud\vec{\lambda}\,\Big\{\prod_{k<l}i(\lambda_{l}-\lambda_{k})\int_{0}^{\infty}\ud u_{kl}\,e^{-iu_{kl}(\lambda_{l}+\lambda_{k})-\varepsilon u_{kl}}\Big\}\quad.\label{e:ST2}
\end{align}

\section{Factorization of the partition function\label{sec:fact}}
In the previous paragraph it was shown that all relevant contributions to the partition function are contained in
\begin{align}
&\hspace{-5mm}\mathpzc{Z}[0]=\frac{(-i\pi)^{\binom{N}{2}}}{\Delta(e_{1},\ldots,e_{N})}\int\ud\vec{\lambda}\,\Delta(\lambda_{1},\ldots,\lambda_{N})\exp[-\sum_{j}e_{j}\lambda_{j}^{2}-g\sum_{j}\lambda_{j}^{4}]\nonumber\\
&\times\Big(\prod_{k<l}\int_{0}^{\infty}\ud u_{kl}\,e^{-iu_{kl}(\lambda_{k}+\lambda_{l})-\varepsilon u_{kl}}\Big)\quad.\label{e:zld1}
\end{align}
In (\ref{e:zld1}) the integrand's dependence on the $u_{kl}$'s is in fact a dependence on 
\begin{equation}
u_{k}=\sum_{j=1}^{k}u_{jk}+\sum_{j=k+1}^{N}u_{kj}\quad.\label{e:uj}
\end{equation}

Substituting the integration variables yields
\begin{align*}
&\hspace{-5mm}\prod_{k<l}\int_{0}^{\infty}\!\ud u_{kl}\, \exp[-i\sum_{j}u_{j}\lambda_{j}-\frac{\varepsilon}{2}\sum_{j}u_{j}]\\
&=\frac{1}{2}\Big(\!\prod_{m=1}^{N}\!\int_{0}^{\infty}\!\!\!\ud u_{m}\Big)\,V_{N}(\vec{u})\exp[-i\sum_{j}u_{j}\lambda_{j}-\frac{\varepsilon}{2}\sum_{j}u_{j}]\quad.
\end{align*}
To find out what the function $V_{N}(\vec{u})$ is, the equation (\ref{e:uj}) is written explicitly as a matrix equation
\begin{equation*}
\left(\begin{array}{cccc}0&u_{12}&\ldots&u_{1N}\\u_{12}&0&\ldots&u_{2N}\\\vdots&\vdots&\ddots&\vdots\\u_{1N}&u_{2N}&\ldots&0\end{array}\right)\left(\begin{array}{c}1\\1\\\vdots\\1\end{array}\right)=\left(\begin{array}{c}u_{1}\\u_{2}\\\vdots\\u_{N}\end{array}\right)\quad.
\end{equation*}
Choosing for the sake of the argument the formulation with all $u_{j}\in(0,1)$, it follows that $h_{j}=1-u_{j}$ lies between $0$ and $1$ and that to such a matrix equation a unique symmetric stochastic matrix 
\begin{equation*}
\left(\begin{array}{cccc}h_{1}&u_{12}&\ldots&u_{1N}\\u_{12}&h_{2}&\ldots&u_{2N}\\\vdots&\vdots&\ddots&\vdots\\u_{1N}&u_{2N}&\ldots&h_{N}\end{array}\right)
\end{equation*}
corresponds. A matrix is stochastic, when all its entries are nonnegative and every row sums to $1$.\\
This implies that the function $V_{N}(\vec{u})$ with all $u_{j}\in[0,1]$ is the volume of the space of symmetric stochastic matrices with diagonal entries $\{1-u_{1},\ldots,1-u_{N}\}$. It is straightforward to check that this space is convex, so that this space is a $\frac{N(N-3)}{2}$-dimensional polytope.\\

The Jacobian of this transformation is $2$. It is not difficult to see that this holds for any $N$. An example of this is the $N=4$-Jacobian in Table~\ref{t:jac4}. For this substitution of variables the determinant is easily calculated. Because these vectors are not perpendicular, we cannot integrate them independently.\\

\footnotesize
\begin{table}[!h]
\hspace*{0cm}\begin{tabular}{l|cccccc}
 & $u_{12}$ & $u_{13}$&$u_{23}$ &$u_{14}$ &$u_{24}$ &$u_{34}$ \\\hline
$u_{1}$  &$1 $&$1 $&$0 $&$1 $&$0 $&$0 $\\
$u_{2}$  &$1 $&$0 $&$1 $&$0 $&$1 $&$0 $\\
$u_{3}$  &$0 $&$1 $&$1 $&$0 $&$0 $&$1 $\\
$u_{4}$  &$0 $&$0 $&$0 $&$1 $&$1 $&$1 $\\
$u_{24}$ &$0 $&$0 $&$0 $&$0 $&$1 $&$0 $\\
$u_{34}$ &$0 $&$0 $&$0 $&$0 $&$0 $&$1 $
\end{tabular}
\caption{The Jacobian corresponding to (\ref{e:uj}) for $N=4$. \label{t:jac4}}
\end{table}
\vspace{5mm}\normalsize

These steps demonstrate that the partition function (\ref{e:zld1}) can be rewritten as
\begin{align}
&\hspace{-5mm}\mathpzc{Z}[0]=\frac{(-i\pi)^{\binom{N}{2}}}{2\Delta(e_{1},\ldots,e_{N})}\int_{0}^{\infty}\!\!\!\!\!\!\ud\vec{u}\,V_{N}(\vec{u})\nonumber\\
&\times\int\ud\vec{\lambda}\,\Delta(\lambda_{1},\ldots,\lambda_{N})\exp[-\sum_{j}(g\lambda_{j}^{4}+e_{j}\lambda_{j}^{2}+iu_{j}\lambda_{j}+\frac{\varepsilon}{2}u_{j})]\label{e:zld2}\quad,
\end{align}
where we have used that the volume of the polytope obeys the scaling law
\begin{equation*}
V_{N}(\vec{u})=M^{\frac{-N(N-3)}{2}}V_{N}(M\vec{u})\qquad\text{for}\quad M\in\mathbb{R}_{+}\quad.
\end{equation*}

In~\cite{jins1} it is shown that the volume of this diagonal subpolytope of symmetric stochastic matrices is given by
\begin{align*}
&\hspace{-5mm}\vol(P_{N}(\vec{h}))=\sqrt{2}e^{\frac{7}{6}}\big(\frac{e(N-\chi)}{N(N-1)}\big)^{\binom{N}{2}}\big(\frac{N(N-1)^{2}}{2\pi(N-\chi)^{2}}\big)^{\frac{N}{2}}\\
&\times\exp[-\frac{(N\!-\!1)^{2}}{2(N\!-\!\chi)^{2}}(N\!+\!2)\sum_{j}(h_{j}\!-\!\frac{\chi}{N})^{2}]\exp[-\frac{N(N\!-\!1)^{3}}{3(N\!-\!\chi)^{3}}\sum_{j}(h_{j}\!-\!\frac{\chi}{N})^{3}]\\
&\times\exp[-\frac{N(N-1)^{4}}{4(N-\chi)^{4}}\sum_{j}(h_{j}-\frac{\chi}{N})^{4}]\exp[\frac{(N-1)^{4}}{4(N-\chi)^{4}}\big(\sum_{j}(h_{j}-\frac{\chi}{N})^{2}\big)^{2}]\quad,
\end{align*}
provided that for all $j=1,\ldots,N$
\begin{equation*}
\lim_{N\rightarrow\infty}N^{\frac{1}{4}}\frac{N-1}{N-\chi}\cdot\big|h_{j}-\frac{\chi}{N}\big|=0\quad
\end{equation*}
where $\chi=\sum_{j}h_{j}$. Substituting $h_{j}=1-u_{j}$ and $\chi=N-S$ with $S=\sum_{j}u_{j}$ taking values in $[0,N]$ now yields
\begin{align}
&\hspace{-5mm}V_{N}(\vec{u})=\sqrt{2}e^{\frac{7}{6}}\big(\frac{eS}{N(N\!-\!1)}\big)^{\binom{N}{2}}\big(\frac{N(N\!-\!1)^{2}}{2\pi S^{2}}\big)^{\frac{N}{2}}\nonumber\\
&\times\exp[-\frac{(N\!-\!1)^{2}}{2S^{2}}(N\!+\!2)\sum_{j}(u_{j}\!-\!\frac{S}{N})^{2}]\nonumber\\
&\times\exp[\frac{N(N-1)^{3}}{3S^{3}}\sum_{j}(u_{j}-\frac{S}{N})^{3}]\exp[-\frac{N(N-1)^{4}}{4S^{4}}\sum_{j}(u_{j}-\frac{S}{N})^{4}]\nonumber\\
&\times\exp[\frac{(N-1)^{4}}{4S^{4}}\big(\sum_{j}(u_{j}-\frac{S}{N})^{2}\big)^{2}]\quad,\label{e:polvol}
\end{align}
provided that
\begin{equation}
\lim_{N\rightarrow\infty}N^{\frac{1}{4}}\frac{N-1}{S}\cdot\big|u_{j}-\frac{S}{N}\big|=0\quad.\label{e:polvolcond}
\end{equation}
This condition implies that
\begin{equation*}
\lim_{N\rightarrow\infty}\sum_{j=1}^{N}N^{-1+\frac{k}{4}}\big(\frac{N-1}{S}\big)^{k}\cdot\big|u_{j}-\frac{S}{N}\big|^{k}=0\qquad\forall k\geq2\quad.
\end{equation*}
Asymptotically, these conditions cover almost all volume of the polytope of symmetric stochastic matrices. The lion's share of the volume is located at small $\chi$, or large $S$.\\

The polytope volume (\ref{e:polvol}) vanishes as any $u_{j}\rightarrow\infty$, so that the Schwinger trick is regularized by the polytope volume and $\varepsilon$ in (\ref{e:zld2}) may be set to zero.\\

In (\ref{e:zld1}) the partition function is factorized. All eigenvalue integrands are of the same form. The price to pay for this is the doubling of the number of integrals, where it is important that the polytope volume is itself factorized. Apart from a subleading factor, this is the case.

\section{No coupling\label{sec:noco}}
So far, the factorization of the partition function is little more than a nice idea. To test whether this evaluation method has any chance of succeeding we return to the free theory. The partition function for the free model (\ref{e:zerocoup}) is rewritten using the polytope volume. Comparing the outcome to the starting point should provide some information on this matter.\\

Additionally, the various steps up to this point have made the expressions only more complicated. A way to find out how these expressions should be treated is through a test calculation, where the outcome is known in advance. Introducing the polytope volume in (\ref{e:zerocoup}) yields
\begin{align}
&\hspace{-5mm}\lim_{g\rightarrow0}\mathpzc{Z}[0]=\big\{\prod_{k=1}^{N}\sqrt{\frac{\pi}{e_{k}}}\big\}\prod_{k<l}\frac{\pi}{e_{k}+e_{l}}\label{e:tc1}\\
&=\pi^{\binom{N}{2}}\big\{\prod_{k=1}^{N}\sqrt{\frac{\pi}{e_{k}}}\big\}\prod_{k<l}\int_{0}^{\infty}\!\!\ud u_{kl}\,e^{-u_{kl}(e_{k}+e_{l})}\nonumber\\
&=\frac{1}{2}\pi^{\binom{N}{2}}\big\{\prod_{k=1}^{N}\sqrt{\frac{\pi}{e_{k}}}\big\}\int_{0}^{\infty}\!\!\!\!\!\!\ud^{N}\vec{u}\,V_{N}(\vec{u})e^{-\sum_{m}u_{m}e_{m}}\nonumber\\
&=\frac{1}{2}\pi^{\binom{N}{2}}\big\{\!\prod_{k=1}^{N}\!\sqrt{\frac{\pi}{e_{k}}}\big\}(N\!-\!1)\!\!\int_{-\infty}^{\infty}\!\!\!\!\!\!\!\ud\kappa\!\int_{-\infty}^{\infty}\!\!\!\!\!\!\!\!\ud q\!\int_{0}^{\infty}\!\!\!\!\!\!\!\!\ud S\!\int_{0}^{\sqrt{N}}\!\!\!\!\!\!\!\!\ud Q\!\int_{-N^\frac{1}{4}}^{N^\frac{1}{4}}\!\!\ud\vec{x}\,\frac{\sqrt{2N}e^{7/6}}{(2\pi)^{N/2}S}\nonumber\\
&\times\exp[-2\pi i\kappa\!\sum_{j}\!x_{j}-\!\sum_{j}\!e_{j}S(\frac{x_{j}}{\sqrt{N}}\!+\!\frac{N\!-\!1}{N})]\big(\frac{eS}{N}\big)^{\binom{N}{2}}\exp[-\frac{N\!+\!2}{2N}\!\sum_{j}\!x_{j}^{2}]\nonumber\\
&\times\exp[\frac{\sum_{j}x_{j}^{3}}{3\sqrt{N}}-\frac{\sum_{j}x_{j}^{4}}{4N}+\frac{Q^{2}}{4}]\exp[2\pi iq(Q-\sum_{j}\frac{x_{j}^{2}}{N})]\quad,\label{e:tc3}
\end{align}
where in the last step the polytope volume (\ref{e:polvol}) was subtituted and the transformations $u_{j}\rightarrow x_{j}+S/N$, $S\rightarrow S(N-1)$, $x_{j}\rightarrow x_{j}S/\sqrt{N}$, $Q\rightarrow QS^{2}$, $q\rightarrow qS^{-2}$ and $\kappa\rightarrow \kappa \sqrt{N}/S$ performed respectively. The application range $|u_{j}-S/N|\ll SN^{-\frac{5}{4}}$ of the polytope volume formula (\ref{e:polvolcond}) implies integration boundaries for the integration parameters $x_{j}$ in (\ref{e:tc3}).\\

It is not straightforward to see what the most convenient integration order in (\ref{e:tc3}) is. What can be seen is that the integral over $S$ can be performed directly. This yields a Gamma function and a fraction depending on $\sum_{j}x_{j}e_{j}$ to the power $\small{\binom{N}{2}}$. For the integral over $x_{j}$ it becomes necessary to make some assumptions on the kinetic model parameters $e_{j}$. Assuming that
\begin{equation}
e_{j}=\xi(1+\tilde{\varepsilon}_{j})\qquad\text{, where }\xi=\frac{1}{N}\sum_{j=1}^{N}e_{j}\label{e:kinpar}
\end{equation}
with $|\tilde{\varepsilon}_{j}|\ll N^{-\frac{1}{4}}$ allows us to approximate the $x_{j}$-dependence in the fraction $\sum(j)e_{j}(\frac{N-1}{N}+\frac{x_{j}}{\sqrt{N}})$ with exponentials, which can be integrated by the stationary phase method. This is also used for the integral over $\kappa$. The remaining integrals are Fourier transforms of the Dirac delta and are therefore straightforward to integrate. All details of this computation can be found in~\cite{jins2}. 

By application of Stirling's formula (\ref{e:gammastir}) to the Gamma function the partition function of the free theory computed via the polytope volume
\begin{align}
&\hspace{-5mm}\lim_{g\rightarrow0}\mathpzc{Z}[0]=\big(\frac{\pi}{2\xi}\big)^{\binom{N}{2}}\big\{\!\prod_{k=1}^{N}\!\sqrt{\frac{\pi}{e_{k}}}\big\}\exp[\frac{N\!-\!2}{8}\sum_{j}\!\tilde{\varepsilon}_{j}^{2}\!-\!\frac{N\!-\!6}{24}\sum_{j}\!\tilde{\varepsilon}_{j}^{3}\!+\!\frac{N}{64}\sum_{j}\!\tilde{\varepsilon}_{j}^{4}]\nonumber\\
&\times\exp[\frac{3}{64}(\sum_{j}\tilde{\varepsilon}_{j}^{2})^{2}-\frac{1}{16}(\sum_{j}\tilde{\varepsilon}_{j}^{2})(\sum_{j}\tilde{\varepsilon}_{j}^{3})+\frac{7}{128}(\sum_{j}\tilde{\varepsilon}_{j}^{2})(\sum_{j}\tilde{\varepsilon}_{j}^{4})]\nonumber\\
&\times\exp[\frac{3}{128}(\sum_{j}\tilde{\varepsilon}_{j}^{3})^{2}-\frac{5}{128}(\sum_{j}\tilde{\varepsilon}_{j}^{3})(\sum_{j}\tilde{\varepsilon}_{j}^{4})+\frac{1}{16N}(\sum_{j}\tilde{\varepsilon}_{j}^{2})^{3}]\nonumber\\
&\exp[-\frac{11}{128N}(\sum_{j}\tilde{\varepsilon}_{j}^{2})^{2}(\sum_{j}\tilde{\varepsilon}_{j}^{3})]\label{e:tc6d}\quad
\end{align}
is obtained. This is to be compared to
\begin{align}
&\hspace{-5mm}\lim_{g\rightarrow0}\mathpzc{Z}[0]=\big\{\prod_{k=1}^{N}\sqrt{\frac{\pi}{e_{k}}}\big\}\prod_{k<l}\frac{\pi}{e_{k}+e_{l}}=\big\{\prod_{k=1}^{N}\sqrt{\frac{\pi}{e_{k}}}\big\}\big(\frac{\pi}{2\xi}\big)^{\binom{N}{2}}\nonumber\\
&\times\exp\big[\!-\!\sum_{k<l}\!\big\{\!\big(\frac{\tilde{\varepsilon}_{k}\!+\!\tilde{\varepsilon}_{l}}{2}\big)-\frac{1}{2}\big(\frac{\tilde{\varepsilon}_{k}\!+\!\tilde{\varepsilon}_{l}}{2}\big)^{2}+\frac{1}{3}\big(\frac{\tilde{\varepsilon}_{k}\!+\!\tilde{\varepsilon}_{l}}{2}\big)^{3}-\frac{1}{4}\big(\frac{\tilde{\varepsilon}_{k}\!+\!\tilde{\varepsilon}_{l}}{2}\big)^{4}\big\}\big]\nonumber\\
&=\big\{\prod_{k=1}^{N}\sqrt{\frac{\pi}{e_{k}}}\big\}\big(\frac{\pi}{2\xi}\big)^{\binom{N}{2}}\exp[\frac{N-2}{8}\sum_{j}\tilde{\varepsilon}_{j}^{2}-\frac{N-4}{24}\sum_{j}\tilde{\varepsilon}_{j}^{3}]\nonumber\\
&\times\exp[\frac{N\!-\!8}{64}\sum_{j}\!\tilde{\varepsilon}_{j}^{4}\!+\!\frac{3}{64}(\sum_{j}\!\tilde{\varepsilon}_{j}^{2})^{2}\!-\!\frac{N\!-\!16}{160}\sum_{j}\!\tilde{\varepsilon}_{j}^{5}\!-\!\frac{1}{16}(\sum_{j}\tilde{\varepsilon}_{j}^{2})(\sum_{j}\tilde{\varepsilon}_{j}^{3})]\nonumber\\
&\times\exp[\frac{N-32}{384}\sum_{j}\tilde{\varepsilon}_{j}^{6}+\frac{5}{128}(\sum_{j}\tilde{\varepsilon}_{j}^{2})(\sum_{j}\tilde{\varepsilon}_{j}^{4})+\frac{5}{96}(\sum_{j}\tilde{\varepsilon}_{j}^{3})^{2}]\quad.\label{e:tc7}
\end{align}
This is the same, provided that $\tilde{\varepsilon}_{j} \ll N^{-\frac{2}{5}}$. Adding an extra term to the polytope volume computation would allow $\tilde{\varepsilon}_{j} \ll N^{-\frac{1}{3}}$.\\

The above computation has not yielded new insight into the free theory's partition function. However, it does show that the method of factorization and integration against the polytope volume functions.\\
However, the polytope volume calculation turned all parameters into symmetric sums, whereas the direct computation is given in pairs of eigenvalues. The difference stems from the asymptotic formulation of the polytope volume, where the matrix structure has disappeared. This leads to a trade-off between structural integrity and computability.

\section{Weak coupling\label{sec:weco}}
Inspired by the success and insights of the previous paragraph one may try to repeat the computation for weak coupling. The computation without coupling in Paragraph~\ref{sec:noco} shows that the factorization procedure with the asymptotic polytope volume alters the partition function's structure. The strictly positive coupling means that the eigenvalue integrals must be performed after factorization.\\

The starting point is (\ref{e:zld2}) with (\ref{e:polvol}), so that
\begin{align}
&\hspace{-5mm}\mathcal{Z}[0]=\frac{\sqrt{2N}e^{\frac{7}{6}}(N-1)}{2\Delta(e_{1},\ldots,e_{N})}\big(\frac{1}{2\pi}\big)^{\frac{N}{2}}\!\!\!\int_{-\infty}^{\infty}\!\!\!\!\!\!\!\ud\kappa\!\int_{0}^{\infty}\frac{\ud S}{S}\!\int_{-\infty}^{\infty}\!\!\!\!\!\!\!\ud\vec{x}\!\int_{-\infty}^{\infty}\!\!\!\!\!\!\!\ud q\!\int_{0}^{\infty}\!\!\!\!\!\!\!\ud Q\!\,\big(\frac{-\pi ieS}{N}\big)^{\binom{N}{2}}\nonumber\\
&\times\!\int_{-\infty}^{\infty}\!\!\!\!\!\!\!\ud\vec{\lambda}\,\Delta(\lambda_{1},\ldots,\lambda_{N})\,\exp[\sum_{j}\big(\!-e_{j}\lambda_{j}^{2}-g\lambda_{j}^{4}-iS\frac{x_{j}}{\sqrt{N}}\lambda_{j}-iS\frac{N\!-\!1}{N}\lambda_{j}\big)]\nonumber\\
&\times \exp[2\pi i q (Q-\sum_{j}\frac{x_{j}^{2}}{N})]\exp[-2\pi i\kappa \sum_{j}x_{j}]\exp[-\frac{N+2}{2N}\sum_{j}x_{j}^{2}]\nonumber\\
&\times\exp[\frac{1}{3\sqrt{N}}\sum_{j}x_{j}^{3}]\exp[\frac{-1}{4N}\sum_{j}x_{j}^{4}]\exp[\frac{Q^{2}}{4}]\label{e:wc0}
\end{align}
is obtained. The same transformations as in Paragraph~\ref{sec:noco} were used to get here. The integration order of the extra integrals is borrowed from Paragraph~\ref{sec:noco}. This means that the integral over $S$ is performed first. Integrating then over $x_{j}$ using the stationary phase method and scaling $\kappa\rightarrow \kappa/\sqrt{N}$ yields
\begin{align}
&\hspace{-5mm}\mathcal{Z}[0]=\frac{\sqrt{2}e^{\frac{7}{6}}(N-1)\Gamma(\binom{N}{2})}{2\Delta(e_{1},\ldots,e_{N})}\int_{-\infty}^{\infty}\!\!\!\!\!\!\!\ud\kappa\!\int_{-\infty}^{\infty}\!\!\!\!\!\!\!\ud q\!\int_{0}^{\infty}\!\!\!\!\!\!\!\ud Q\!\int_{-\infty}^{\infty}\!\!\!\!\!\!\!\ud \mu\!\int_{-\infty}^{\infty}\!\!\!\!\!\!\!\ud X\!\int_{-\infty}^{\infty}\!\!\!\!\!\!\!\ud \nu\!\int_{-\infty}^{\infty}\!\!\!\!\!\!\!\ud \Lambda\,\nonumber\\
&\times\big(\frac{-\pi e}{N(\Lambda +X)}\big)^{\binom{N}{2}}\int_{-\infty}^{\infty}\!\!\!\!\!\!\!\ud\vec{\lambda}\,\Delta(\lambda_{1},\ldots,\lambda_{N})\,\exp[\sum_{j}\big(-e_{j}\lambda_{j}^{2}-g\lambda_{j}^{4}\big)]\nonumber\\
&\times \exp[2\pi i q Q+\frac{Q^{2}}{4}+2\pi i\mu X+2\pi i\nu(\Lambda-\frac{N\!-\!1}{N}\sum_{j}\lambda_{j})]\exp[\frac{5}{6}-\frac{3}{4}]\nonumber\\
&\times(1+\frac{2}{N}+\frac{4\pi iq}{N})^{-\frac{N}{2}}\exp[\sum_{j}\frac{[-2\pi i(\kappa+\mu\lambda_{j})]^{2}}{2N\{1+\frac{2}{N}+\frac{4\pi iq}{N}\}}]\label{e:wc17b}\\
&\times\exp[\sum_{j}\frac{[\ldots]^{3}}{3N^{2}\{\ldots\}^{3}}]\exp[\sum_{j}\frac{[\ldots]}{N\{\ldots\}^{2}}]\exp[-\sum_{j}\frac{[\ldots]^{4}}{4N^{3}\{\ldots\}^{4}}]\nonumber\\
&\times\exp[\sum_{j}\!\frac{[\ldots]^{4}}{2N^{3}\{\ldots\}^{5}}]\exp[\sum_{j}\!\frac{-3[\ldots]^{2}}{2N^{2}\{\ldots\}^{3}}]\exp[\sum_{j}\!\frac{2[\ldots]^{2}}{N^{2}\{\ldots\}^{4}}]\label{e:wc17}\quad.
\end{align}
The expressions inside the brackets $[\ldots]$ and $\{\ldots\}$ are the same as the expressions in the same brackets in the exponential on line (\ref{e:wc17b}). Selecting leading terms and using that $\sum_{j}\lambda_{j}=\frac{N}{N-1}\Lambda$ gives
\begin{align*}
&\hspace{-5mm}\int_{-\infty}^{\infty}\ud\kappa\,\exp[-2\pi i\kappa\big(1-\frac{4}{N}-\frac{8\pi iq}{N}-2\pi i\frac{\mu}{N\!-\!1}\Lambda(1-\frac{2}{N}-\frac{4\pi iq}{N})\big)]\\
&\times\exp[-2\pi^{2}\kappa^{2}(1-\frac{2}{N}-\frac{4\pi iq}{N}-\frac{4\pi i\mu\Lambda}{N(N\!-\!1)})]\exp[\frac{8\pi^{3}i\kappa^{3}}{3N}]\\
&=\frac{1}{\sqrt{2\pi}}\exp[-\frac{1}{2}+\frac{2\pi i\mu\Lambda}{N-1}(1-\frac{4}{N})+\frac{2\pi^{2}\mu^{2}\Lambda^{2}}{(N-1)^{2}}]\quad.
\end{align*}
Repeating these steps for $\mu$ results in
\begin{align*}
&\hspace{-5mm}\int_{-\infty}^{\infty}\ud\mu\,\exp[2\pi i\mu\big(X+\frac{\Lambda}{N-1}(1-\frac{4}{N}-\frac{8\pi iq}{N})-\frac{\Lambda}{N-1}(1-\frac{4}{N}-\frac{8\pi iq}{N})\big)]\\
&\times\exp[2\pi^{2}\mu^{2}\big(\frac{\Lambda^{2}}{(N-1)^{2}}-\frac{4\Lambda^{2}}{N(N-1)^{2}}-\frac{1}{N}\sum_{j}\lambda^{2}\big)]=\delta(X)\quad.
\end{align*}
The integrand $\lambda_{j}^{N}\exp[-e_{j}\lambda_{j}^{2}]$ is maximal for $\tilde{\lambda}_{j}^{2}=N/(2e_{j})$. A large $\xi$ in the parameters convention (\ref{e:kinpar}) ensures small $\tilde{\lambda}$. If the quadratic term in the exponential is small in the integral over $\mu$, this yields then $\delta(X)$. This makes the integral over $X$ trivial. Putting things together yields
\begin{align}
&\hspace{-5mm}\mathcal{Z}[0]=\frac{e^{-\frac{1}{4}}(N-1)\Gamma(\binom{N}{2})}{2\sqrt{\pi}\Delta(e_{1},\ldots,e_{N})}\int_{-\infty}^{\infty}\!\!\!\!\!\!\!\ud q\!\int_{0}^{\infty}\!\!\!\!\!\!\!\ud Q\!\int_{-\infty}^{\infty}\!\!\!\!\!\!\!\ud \nu\!\int_{-\infty}^{\infty}\!\!\!\!\!\!\!\ud \Lambda\,\big(\frac{-\pi e}{N\Lambda}\big)^{\binom{N}{2}}\nonumber\\
&\times\int_{-\infty}^{\infty}\!\!\!\!\!\!\!\ud\vec{\lambda}\,\Delta(\lambda_{1},\ldots,\lambda_{N})\,\exp[\sum_{j}\big(-e_{j}\lambda_{j}^{2}-g\lambda_{j}^{4}\big)]\nonumber\\
&\times \exp[2\pi i q (Q-1)+\frac{Q^{2}}{4}]\exp[2\pi i\nu(\Lambda-\frac{N-1}{N}\sum_{j}\lambda_{j})]\label{e:wc18}\quad
\end{align}
which shows that the integral over $q$ and $Q$ are straightforward too. The determinant is rewritten using
\begin{equation}
\Delta(\lambda_{1},\ldots,\lambda_{N})=\lim_{\delta\rightarrow0}\big(i\delta\big)^{\binom{N}{2}}\sum_{\sigma\in\mathcal{S}_{N}}\sgn(\sigma)\exp[\sum_{j}i\delta\sigma(j)\lambda_{j}]\label{e:detexp}\quad,
\end{equation}
where $\mathcal{S}_{N}$ is the symmetric group of the set of $N$ elements. Integrating over $\lambda_{j}$ yields
\begin{align*}
&\hspace{-5mm}\int_{-\infty}^{\infty}\ud\lambda_{j}\,\exp[-g\lambda_{j}^{4}-e_{j}\lambda_{j}^{2}-\lambda_{j}(\frac{2\pi i\nu(N\!-\!1)}{N}-i\delta\sigma(j))]\\
&=\sqrt{\frac{\pi}{e_{j}}}\exp[-\frac{3g}{4e_{j}^{2}}-\frac{1}{4e_{j}}(\frac{2\pi \nu(N\!-\!1)}{N}-\delta\sigma(j))^{2}]\\
&\times\exp[-\frac{g}{16e_{j}^{4}}(\frac{2\pi \nu(N\!-\!1)}{N}-\delta\sigma(j))^{4}-\frac{3g}{4e_{j}^{3}}(\frac{2\pi \nu(N\!-\!1)}{N}-\delta\sigma(j))^{2}]\\
&\approx\sqrt{\frac{\pi}{e_{j}}}\exp[-\frac{3g}{4e_{j}^{2}}-\frac{\pi^{2}\nu^{2}(N\!-\!1)^{2}}{N^{2}e_{j}}+\frac{\pi\nu(N\!-\!1)\delta\sigma(j)}{Ne_{j}}]\quad.
\end{align*}
Here it is used that the coupling is small, so that the additional terms in the exponential may be ignored. Technically, it is not necessary to do this, although it makes the analysis simpler. Using the same integration strategy once for $\nu$ 
leads to the final integral
\begin{align*}
&\hspace{-5mm}\mathcal{Q}=\int_{-\infty}^{\infty}\ud\Lambda\,\big(i\delta\Lambda\big)^{-\binom{N}{2}}\exp[-\frac{N^{2}\Lambda^{2}}{(N\!-\!1)^{2}(\sum_{m}e_{m}^{-1})}+\sum_{j}\frac{i\delta N\Lambda\sigma(j)}{(N\!-\!1)(\sum_{m}e_{m}^{-1})e_{j}}\\
&+\frac{1}{\sum_{m}e_{m}^{-1}}(\sum_{j}\frac{\delta\sigma(j)}{2e_{j}})^{2}]\\
&=\int_{-\infty}^{\infty}\!\!\!\!\!\!\ud\Lambda\,\big(i\delta\Lambda\big)^{-\binom{N}{2}}\exp[-\frac{N^{2}\Lambda^{2}}{(N\!-\!1)^{2}(\sum_{m}e_{m}^{-1})}\!+\!\sum_{j}\frac{i\delta N\Lambda\sigma(j)}{(N\!-\!1)(\sum_{m}e_{m}^{-1})e_{j}}]\\
&\times\exp[-\frac{(N\!-\!1)^{2}(\sum_{m}e_{m}^{-1})}{4N^{2}\Lambda^{2}}\big(\sum_{j}\frac{i\delta N\Lambda\sigma(j)}{(N\!-\!1)(\sum_{m}e_{m}^{-1})e_{j}}\big)^{2}]\quad.
\end{align*}
In this formulation one may recognise a matrix determinant (\ref{e:detexp}). The first nonvanishing coefficient is that of $\delta^{\binom{N}{2}}$. Applying the power series expansion
\begin{equation*}
\partial_{x=0}^{\binom{N}{2}}e^{x+ax^{2}}=\partial_{x=0}^{\binom{N}{2}}\sum_{m=0}^{\infty}\frac{x^{m}}{m!}\sum_{l=0}^{\lfloor m/2\rfloor}\frac{(m!)\,a^{l}}{(l!)\cdot(m-2l)!}
\end{equation*}
with (\ref{e:detexp}) in the opposite direction yields a Vandermonde-determinant of the reciprocals of the kinetic model parameters. The remaining terms take the form
\begin{align}
&\hspace{-5mm}\mathcal{Q}=\frac{(N\!-\!1)\sqrt{\sum_{m}e_{m}^{-1}}}{N}\big(\frac{N}{(N\!-\!1)(\sum_{m}e_{m}^{-1})}\big)^{\binom{N}{2}}\Delta(\frac{1}{e_{1}},\ldots,\frac{1}{e_{N}})\nonumber\\
&\times\int\ud \Lambda\,\exp[-\Lambda^{2}]\cdot\sum_{l=0}^{\lfloor\frac{N(N\!-\!1)}{4}\rfloor}\frac{\binom{N}{2}!}{(l!)\cdot(\binom{N}{2}-2l)!}\big(\frac{-1}{4\Lambda^{2}}\big)^{l}\label{e:wc19}
\end{align}
The formulation in (\ref{e:wc19}) is ambiguous, because it is not clear how the integral over $\Lambda$ should be performed for $l>0$. As a real integral a single term is divergent. Divergencies corresponding to $\Lambda=0$ are not present in (\ref{e:ZJ0c}), but may be interpreted as the asymptotic generalization of termwise divergencies cancelled by the matrix symmetry.\\

Alternatively, extending the upper bound of the summation to infinity shows this is a Meijer $G$-function, which can be integrated against any other such function. The result of this is the multiplication by a factor $2^{\binom{N}{2}-1}$ of the $l=0$-term. This factor is easily overlooked, when using other integration orders.\\

Including only the $l=0$-term in (\ref{e:wc19}) sets
\begin{align*}
&\mathcal{Q}_{0}=\sqrt{\pi}\frac{(N\!-\!1)\sqrt{\sum_{m}e_{m}^{-1}}}{N}\big(\frac{N}{(N\!-\!1)(\sum_{m}e_{m}^{-1})}\big)^{\binom{N}{2}}\Delta(\frac{1}{e_{1}},\ldots,\frac{1}{e_{N}})\nonumber\quad.
\end{align*}

All integrals are now computed. The Gamma function is approximated by Stirling's formula 
\begin{equation}
\Gamma\Big(\binom{N}{2}\Big)=2\sqrt{\frac{\pi}{ N(N-1)}}\Big(\frac{N(N\!-\!1)}{2e}\Big)^{\binom{N}{2}}\times\big(1+\mathcal{O}(N^{-2})\big)\quad\label{e:gammastir}
\end{equation}
and the Vandermonde determinant of the inverses can be written as
\begin{equation*}
\Delta(\frac{1}{e_{1}},\ldots,\frac{1}{e_{N}})=\prod_{k<l}\frac{1}{e_{l}}-\frac{1}{e_{k}}=(-1)^{\binom{N}{2}}\Delta(e_{1},\ldots,e_{N})\prod_{m=1}^{N}e_{m}^{1-N}\quad.
\end{equation*}
Putting this together yields the partition function for small but strictly positive coupling
\begin{align}
&\hspace{-5mm}\mathcal{Z}[0]=\sqrt{\frac{N\!-\!1}{N}}\big[\!\prod_{m=1}^{N}\!\sqrt{\frac{\pi}{e_{m}}}e_{m}^{1-N}\big]\big(\frac{\pi N}{2(\sum_{m}e_{m}^{-1})}\big)^{\binom{N}{2}}\exp[-\sum_{m}\frac{3g}{4e_{m}^{2}}]\label{e:wc20}\;.
\end{align}
To compare this to the result without coupling (\ref{e:tc7}) the parameter convention
\begin{equation*}
\xi=\frac{1}{N}\sum_{j}e_{j}\qquad\text{and}\qquad e_{j}=\xi(1+\tilde{\varepsilon}_{j})
\end{equation*}
for small $\tilde{\varepsilon}_{j}\ll 1$ is used again. This yields
\begin{align}
&\hspace{-5mm}\mathcal{Z}[0]=\big[\!\prod_{m=1}^{N}\!\sqrt{\frac{\pi}{e_{m}}}\big]\big(\frac{\pi}{2\xi}\big)^{\binom{N}{2}}\exp[-\sum_{m}\frac{3g}{4e_{m}^{2}}]\exp[-\binom{N}{2}\log\big(\frac{1}{N}\sum_{m}\frac{1}{1+\tilde{\varepsilon}_{m}}\big)]\nonumber\\
&\times\exp[-(N\!-\!1)\sum_{m}\log(1+\tilde{\varepsilon}_{m})]\nonumber\\
&=\big[\!\prod_{m=1}^{N}\!\sqrt{\frac{\pi}{e_{m}}}\big]\big(\frac{\pi}{2\xi}\big)^{\binom{N}{2}}\exp[-\sum_{m}\frac{3g}{4e_{m}^{2}}]\exp[\frac{N\!-\!1}{6}\sum_{m}\tilde{\varepsilon}_{m}^{3}-\frac{N\!-\!1}{4}\sum_{m}\tilde{\varepsilon}_{m}^{4}]\nonumber\\
&\times\exp[\frac{3(N\!-\!1)}{10}\sum_{m}\tilde{\varepsilon}_{m}^{5}-\frac{N\!-\!1}{3}\sum_{m}\tilde{\varepsilon}_{m}^{6}]\nonumber\\
&\times\exp[\frac{N\!-\!1}{4N}(\sum_{m}\tilde{\varepsilon}_{m}^{2})^{2}-\frac{1}{2}(\sum_{m}\tilde{\varepsilon}_{m}^{2})(\sum_{n}\tilde{\varepsilon}_{n}^{3})+\frac{1}{2}(\sum_{m}\tilde{\varepsilon}_{m}^{2})(\sum_{n}\tilde{\varepsilon}_{n}^{4})]\nonumber\\
&\times\exp[\frac{1}{4}(\sum_{m}\tilde{\varepsilon}_{m}^{3})^{2}-\frac{1}{6N}(\sum_{m}\tilde{\varepsilon}_{m}^{2})^{3}]\quad.\label{e:wc21}
\end{align}

The disappointing conclusion is that there is no neighbourhood of parameters such that this is equal to (\ref{e:tc6d}) as one would hope. Any deviation from the symmetric situation $e_{m}=\xi$ modifies the partition function significantly. Although the integration against the polytope volume allows an evaluation of the partition function that is nonpertubative in the coupling, the kinetic parameters are fixed to the symmetric case. Also this is reminiscent of perturbative quantum field theory. To make connection to the free theory, some model parameters must be fixed to their trivial values.\\

The matrix structure in the regulated partition function has been removed by the polytope volume to factorize the computation. This was described as a trade-off between structural integrity and computability. Continuing with this new structure instead of the (reformulated) matrices may cause artificial structures to appear. However, the numerical proximity demands that a limit case must exist, in which the original value is retrieved. The difference between the obtained partition function and this limit case is either vanishing or diverging. The latter corresponds to the subtraction of divergent terms to obtain the desired partition function. This practice is common in perturbative quantum field theory, where it is performed on the level of Feynman diagrams. 

\section{Conclusions\label{sec:conc}}
The matrix basis for the Moyal plane connects \textsc{qft} to Hermitean matrix models. Except for cubic theories, the intertwining of the matrix eigenvalues forms a large obstacle for the direct evaluation of such models. In this paper a method to overcome this difficulty is discussed. The partition function is factorized through integration against the asymptotic volume of the diagonal subpolytope of symmetric stochastic matrices, so that direct evaluation before renormalization is within reach. Although this process modifies the underlying structure of the model, this is in itself not fatal. The partition function without coupling is rewritten using the polytope volume for kinetic eigenvalues that differ only slightly.\\

However, the asymptotic nature of the polytope volume poses a risk. It is difficult to tell in advance whether the modification of the underlying structure will lead to artificial divergences and analytical changes. This is seen explicitly for the case of weak coupling, where both divergences and analytical changes in the dependence on the kinetic eigenvalues appear.\\

To this end it would be interesting to consider the case of strong coupling of the real scalar quartic \textsc{qft}. The quartic interaction is then the dominant factor and it is natural to assume the kinetic eigenvalues small, which is closer to the technical demands obtained in the regime of weak coupling.

\subsection*{Acknowledgments}
This work was supported by the Deutsche Forschungsgemeinschaft (SFB 878).

\nocite{*}
\bibliography{JMP_JdJ_NpEPfSqqftMpWc}

\end{document}